\documentclass[prc,twocolumn,amsmath,amssymb,superscriptaddress,showpacs]{revtex4}
\usepackage{color,graphics,epsfig}

\newcommand {\snn}      {\sqrt{s_{_{\rm NN}}}}

\newcommand {\pt}       {\ensuremath{p_T}}
\newcommand {\mpt}      {\ensuremath{\langle p_{T} \rangle}}

\newcommand {\deta}     {\ensuremath{\Delta\eta}}
\newcommand {\dphi}     {\ensuremath{\Delta\phi}}

\newcommand{\pTtrig}{\ensuremath{p_{T}^{trig}}}
\newcommand{\pTassoc}{\ensuremath{p_{T}^{assoc}}}
\newcommand{\pTtb}{\ensuremath{<p_{T}^{trig}<}}
\newcommand{\pTab}{\ensuremath{<p_{T}^{assoc}<}}
\newcommand{\GeVc}{{\rm GeV/}$c$}

\begin{document}

\title{Azimuthal di-hadron correlations in d+Au and Au+Au collisions at $\sqrt{s_{NN}}=200$ GeV from STAR}

\affiliation{Argonne National Laboratory, Argonne, Illinois 60439, USA}
\affiliation{University of Birmingham, Birmingham, United Kingdom}
\affiliation{Brookhaven National Laboratory, Upton, New York 11973, USA}
\affiliation{University of California, Berkeley, California 94720, USA}
\affiliation{University of California, Davis, California 95616, USA}
\affiliation{University of California, Los Angeles, California 90095, USA}
\affiliation{Universidade Estadual de Campinas, Sao Paulo, Brazil}
\affiliation{University of Illinois at Chicago, Chicago, Illinois 60607, USA}
\affiliation{Creighton University, Omaha, Nebraska 68178, USA}
\affiliation{Czech Technical University in Prague, FNSPE, Prague, 115 19, Czech Republic}
\affiliation{Nuclear Physics Institute AS CR, 250 68 \v{R}e\v{z}/Prague, Czech Republic}
\affiliation{University of Frankfurt, Frankfurt, Germany}
\affiliation{Institute of Physics, Bhubaneswar 751005, India}
\affiliation{Indian Institute of Technology, Mumbai, India}
\affiliation{Indiana University, Bloomington, Indiana 47408, USA}
\affiliation{Alikhanov Institute for Theoretical and Experimental Physics, Moscow, Russia}
\affiliation{University of Jammu, Jammu 180001, India}
\affiliation{Joint Institute for Nuclear Research, Dubna, 141 980, Russia}
\affiliation{Kent State University, Kent, Ohio 44242, USA}
\affiliation{University of Kentucky, Lexington, Kentucky, 40506-0055, USA}
\affiliation{Institute of Modern Physics, Lanzhou, China}
\affiliation{Lawrence Berkeley National Laboratory, Berkeley, California 94720, USA}
\affiliation{Massachusetts Institute of Technology, Cambridge, MA 02139-4307, USA}
\affiliation{Max-Planck-Institut f\"ur Physik, Munich, Germany}
\affiliation{Michigan State University, East Lansing, Michigan 48824, USA}
\affiliation{Moscow Engineering Physics Institute, Moscow Russia}
\affiliation{City College of New York, New York City, New York 10031, USA}
\affiliation{NIKHEF and Utrecht University, Amsterdam, The Netherlands}
\affiliation{Ohio State University, Columbus, Ohio 43210, USA}
\affiliation{Old Dominion University, Norfolk, VA, 23529, USA}
\affiliation{Panjab University, Chandigarh 160014, India}
\affiliation{Pennsylvania State University, University Park, Pennsylvania 16802, USA}
\affiliation{Institute of High Energy Physics, Protvino, Russia}
\affiliation{Purdue University, West Lafayette, Indiana 47907, USA}
\affiliation{Pusan National University, Pusan, Republic of Korea}
\affiliation{University of Rajasthan, Jaipur 302004, India}
\affiliation{Rice University, Houston, Texas 77251, USA}
\affiliation{Universidade de Sao Paulo, Sao Paulo, Brazil}
\affiliation{University of Science \& Technology of China, Hefei 230026, China}
\affiliation{Shandong University, Jinan, Shandong 250100, China}
\affiliation{Shanghai Institute of Applied Physics, Shanghai 201800, China}
\affiliation{SUBATECH, Nantes, France}
\affiliation{Texas A\&M University, College Station, Texas 77843, USA}
\affiliation{University of Texas, Austin, Texas 78712, USA}
\affiliation{Tsinghua University, Beijing 100084, China}
\affiliation{United States Naval Academy, Annapolis, MD 21402, USA}
\affiliation{Valparaiso University, Valparaiso, Indiana 46383, USA}
\affiliation{Variable Energy Cyclotron Centre, Kolkata 700064, India}
\affiliation{Warsaw University of Technology, Warsaw, Poland}
\affiliation{University of Washington, Seattle, Washington 98195, USA}
\affiliation{Wayne State University, Detroit, Michigan 48201, USA}
\affiliation{Institute of Particle Physics, CCNU (HZNU), Wuhan 430079, China}
\affiliation{Yale University, New Haven, Connecticut 06520, USA}
\affiliation{University of Zagreb, Zagreb, HR-10002, Croatia}

\author{M.~M.~Aggarwal}\affiliation{Panjab University, Chandigarh 160014, India}
\author{Z.~Ahammed}\affiliation{Lawrence Berkeley National Laboratory, Berkeley, California 94720, USA}
\author{A.~V.~Alakhverdyants}\affiliation{Joint Institute for Nuclear Research, Dubna, 141 980, Russia}
\author{I.~Alekseev~~}\affiliation{Alikhanov Institute for Theoretical and Experimental Physics, Moscow, Russia}
\author{J.~Alford}\affiliation{Kent State University, Kent, Ohio 44242, USA}
\author{B.~D.~Anderson}\affiliation{Kent State University, Kent, Ohio 44242, USA}
\author{D.~Arkhipkin}\affiliation{Brookhaven National Laboratory, Upton, New York 11973, USA}
\author{G.~S.~Averichev}\affiliation{Joint Institute for Nuclear Research, Dubna, 141 980, Russia}
\author{J.~Balewski}\affiliation{Massachusetts Institute of Technology, Cambridge, MA 02139-4307, USA}
\author{L.~S.~Barnby}\affiliation{University of Birmingham, Birmingham, United Kingdom}
% ...

\collaboration{STAR Collaboration}\noaffiliation

\begin{abstract}
Yields, correlation shapes, and mean transverse momenta \pt{} of
charged particles associated with intermediate to
high-\pt{} trigger particles ($2.5 < \pt < 10$ \GeVc) in d+Au and
Au+Au collisions at $\snn=200$ GeV are presented. For associated
particles at higher $\pt \gtrsim 2.5$ \GeVc, narrow correlation peaks
are seen in d+Au and Au+Au, indicating that the main production
mechanism is jet fragmentation. At lower associated particle $\pt < 2$
\GeVc, a large enhancement of the near- ($\dphi \sim 0$) and away-side
($\dphi \sim \pi$) associated yields is found, together with a strong
broadening of the away-side azimuthal distributions in Au+Au collisions 
compared to
d+Au measurements, suggesting that other particle production
mechanisms play a role. This is further supported by
the observed significant softening of the away-side associated
particle yield distribution at $\dphi \sim \pi$ in central Au+Au
collisions.
\end{abstract}


\maketitle

\section{Introduction}
The goal of ultra-relativistic heavy-ion collisions is to create a
system of deconfined quarks and gluons at high temperature and density
and study its properties. In the initial stage of the collision, hard
scatterings between partons in the incoming nuclei produce high
transverse momentum (\pt) partons that fragment into jets of hadrons
with a clear back-to-back di-jet signature~\cite{Adams:2003im}.  In
Au+Au collisions, hard partons traverse the hot and dense colored
medium, thus probing the medium through energy
loss~\cite{Gyulassy:1990ye,Wang:1991xy,Baier:1996kr}.

In-medium jet energy loss was first observed at the Relativistic Heavy
Ion Collider (RHIC) as a suppression of hadron spectra at high $\pt$
\cite{Adler:2002xw,Adcox:2001jp} in Au+Au collisions with respect to
p+p collisions. The jet-like structure of hadron production at high
\pt{} was confirmed by measurements of the azimuthal angle difference
\dphi{} distributions of {\it associated} particles in a certain range
of $\pt$ with respect to a {\it trigger} hadron at a higher \pt{}
\cite{Adams:2003im}. At the highest \pt, a suppression of the
away-side yield (around $\dphi \sim \pi$ with respect to the trigger
particle) by a factor 3--5 is observed \cite{Adams:2006yt}. This
suppression is consistent with theoretical calculations that
incorporate in-medium energy loss
\cite{Loizides:2006cs,Zhang:2007ja}. At lower \pt{} of the associated
particles, a strongly broadened away-side structure is seen in Au+Au
collisions, and the associated yields on both the near-side ($\dphi \sim
0$) and away-side ($\dphi \sim \pi$) are enhanced
\cite{Adams:2005ph,:2008cqb}. A number of possible explanations of
the away-side broadening at intermediate \pt{} have been put forward,
ranging from fragmentation products of radiated gluons
\cite{Vitev:2005yg,Polosa:2006hb} to medium response and the
possibility of a
Mach-cone shock wave
\cite{Stoecker:2004qu,CasalderreySolana:2004qm,Renk:2005si,Ruppert:2005uz}.

The \pTtrig\ range used in previous
studies~\cite{Adams:2005ph,:2008cqb} is the region where the
p/$\pi$ ratio is large. The large baryon/meson ratio has been
interpreted as being due to coalescence/recombination of quarks, which 
could also have an impact on the jet-like
correlation yields, especially for trigger particles in the $p_T$ range 2.0 to
4.0 \GeVc{} where 
coalescence/recombination
products~\cite{Hwa:2002zu,Fries:2003vb,Greco:2003xt} may be present.

In this paper, we present a systematic exploration of the azimuthal
di-hadron correlation shapes and yields with centrality and \pt{} of
the trigger (\pTtrig) and associated hadrons (\pTassoc), to
investigate the change from broadened correlation peaks with enhanced
yields at low \pt{} to suppressed away-side yields at high \pt. The
analysis is performed on the large statistics sample of Au+Au
collisions at $\sqrt{s_{NN}}=200$ GeV collected by the STAR experiment
in the RHIC run in 2004. The d+Au data sample from the year 2003 is
used as a reference where no hot and dense matter is formed, because
the minimum bias p+p data collected by STAR has limited
statistics. Earlier studies have shown that di-hadron correlations in
p+p and d+Au collisions are similar \cite{Adams:2003im}.

\section{Experimental setup and data sets}
The measurements presented in this paper were performed with the STAR
detector at RHIC \cite{Ackermann:2002ad}. Charged tracks are
reconstructed with the Time Projection Chamber (TPC) \cite{Anderson2003659}.

For Au+Au collisions, two different online event selections
(triggers), minimum bias and central, were used.  The central trigger
selection was based on the energy deposited in the two Zero-Degree
Calorimeters (ZDCs) which measure spectator fragments and small-angle
particle production \cite{Ackermann:2002ad}. The trigger selected the
most central 12\% of the total hadronic cross section,
based on a maximum energy deposited in the ZDCs and a
minimum multiplicity in the Central Trigger Barrel
(CTB)~\cite{Bieser:2002ah}. The central trigger also uses time
information from the Beam-Beam Counters (BBC) to restrict the primary
vertex position $z_{vtx}$ to be within approximately $\pm 30 $ cm of
the center of the detector along the beam direction. The minimum bias
(MB) trigger is based on a ZDC coincidence (a threshold amount of
energy in each ZDC) and required a minimum multiplicity in the CTB
to reject non-hadronic interactions. For the minimum
bias sample, events were selected to have $|z_{vtx}| < 25$ cm. A total
of 21M minimum bias events and 18M central triggered Au+Au events were
used.

For d+Au collisions, the minimum bias trigger was defined by requiring
that at least one beam-rapidity neutron impinge on the ZDC in the Au
beam direction. The measured minimum bias cross section amounts to
$95 \pm 3\%$ of the total d+Au geometric cross
section~\cite{Adams:2003im}. For d+Au events, the distribution of
primary vertices along the beamline was wider than during the Au+Au
run. The events were selected to be within $\pm 50$
cm from the center of the detector along the beam line. A total of
3.4M d+Au events were selected for this analysis.

The Au+Au events are further divided into centrality classes based on
the uncorrected charged particle multiplicity in the range
$|\eta|<0.5$ as measured by the TPC. We present results for the
following centrality ranges: 0-12\% (from the central triggered data
set), 20-40\%, 40-60\%, and 60-80\% (from the MB data set) of the
total hadronic cross section, with 0\% referring to the most central
collisions.

\section{Data analysis}
Di-hadron correlations are constructed using charged particles
measured in the TPC. All particles are selected to have pseudo-rapidity in the
range $-1.0 < \eta < 1.0$, so that they fall well within the TPC acceptance. To reject background tracks at high \pt,
tracks were required to have at least 20 measured points in the TPC
(out of 45) and a distance of closest approach (dca) to the event
vertex of less than 1 cm to reduce the contribution from secondary particles.

The results are corrected for single particle acceptance and detection
efficiency as well as for the pair acceptance as a function of
$\dphi$. The single particle reconstruction efficiency as a function
of $\eta$, \pt, and centrality is determined using hits from a Monte\
Carlo simulation which are embedded into real data events. The
tracking efficiency depends sensitively on the gain in the
proportional readout chamber of the TPC and thus on the atmospheric
pressure. Uncertainties in details of these effects give rise to an
overall 5\% systematic uncertainty in the absolute yields given in
this paper. In most cases, the uncertainty from the background
subtraction as described in the next section is larger than the
systematic uncertainty from the tracking efficiency. The TPC sector
boundaries introduce a dependence of the pair acceptance on angle
difference $\dphi$, which was determined from mixed events. No
$\Delta\eta$ pair acceptance correction has been
applied. A small inefficiency due to tracks crossing
inside the TPC volume affects the asssociated hadron distribution at
small pair separation in ($\Delta\eta,\Delta\phi$). This effect
manifests itself as a reduced efficiency for small but finite \dphi{},
at positive or negative \dphi, depending on the sign of the curvature
of the associate track. A correction was performed by first
curvature-sorting the distributions and then reflecting a few bins
from the unaffected area to the area where the inefficiency occurs,
thus restoring the symmetry between positive and negative
$\Delta\phi$.

\begin{figure}[hbt!]
\includegraphics[width=\columnwidth]{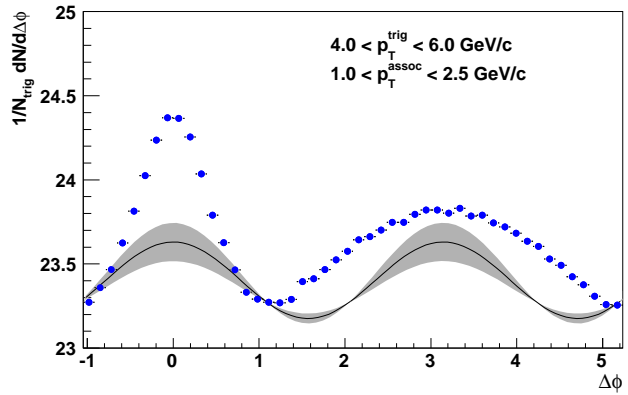}
\caption{\label{fig:dphi_bkg}(Color online) Azimuthal distribution of
  associated charged particles with 1.0 \pTab\ 2.5~\GeVc{} with
  respect to trigger particles with 4.0 \pTtb\ 6.0~\GeVc{} in 0-12\%
  central Au+Au collisions. The curve shows the modulation of the
  background due to elliptic flow $v_2$ and the grey band indicates
  the uncertainty on the elleptic flow of the background (see text).}
\end{figure}

Figure~\ref{fig:dphi_bkg} shows an example azimuthal angle difference
distribution for trigger particles with $4.0 < \pTtrig < 6.0$ GeV/$c$
and associated particles with $1.0 < \pTassoc < 2.5$ GeV/$c$ in 0-12\%
central Au+Au collisions. The distribution is divided by the number of
trigger hadrons to give the associated yield per trigger hadron. The
associated hadron distribution contains a background of uncorrelated
particles which has a $\cos{(2\Delta\phi)}$ modulation due to the
correlation of all particles with the reaction plane through elliptic
flow, $v_2$. We model the background using the function $B(1+2\langle
v_2^{trig}\rangle\langle v_2^{assoc}\rangle\cos{(2\Delta\phi}))$,
where the $v_2$ values are from separate flow measurements
\cite{Adams:2004bi}. The function is normalized to the data in the
region $0.8<|\dphi|<1.2$, where the signal is apparently small, and
then subtracted. This background normalisation
procedure is often referred to as the ZYA1 ('Zero Yield at 1 radian')
or ZYAM ('Zero Yield At Minimum') method \cite{Ajitanand:2005jj}.

The method of normalising the combinatorial
background level in the region around $|\dphi|=1$ has first been used
for di-hadron correlations at higher
momenta~\cite{Adams:2005ph,Adams:2006yt}, where there are narrow peaks
on the near- and away-side, separated by a largely `signal-free'
region. At lower \pt, the correlation peaks are broader, and there is
no clear signal-free region, so that the background normalisation is
more ambiguous. In addition, due to the larger combinatorial
background, the elliptic flow modulation of the background is of
similar size as the trigger-associated-hadron correlation signal. The
ZYA1 method provides a simple prescription to separate signal and
background, which we will use throughout the paper. An alternative
approach would be to decompose the correlation shape using a fit
function with components representing the flow modulation of the
background and the assumed shapes of near- and away-side correlation peaks \cite{Adams:2006tj,Trainor:2009gj}. The unsubtracted azimuthal hadron
distributions are provided in the Appendix and can be used for
such a procedure.

The nominal $v_2$ value used for the subtraction is the mean of the
$v_2$ measured using the reaction plane method with the Forward TPC and
the four-particle cumulant method, which have different sensitivity to
non-flow effects and flow fluctuations~\cite{Adams:2004bi} (line in
Fig.~\ref{fig:dphi_bkg}). The difference between the two results is
used as the estimate of the systematic uncertainty in $v_2$ and this
range is shown by the band in Fig.~\ref{fig:dphi_bkg}. For the d+Au
results a constant pedestal (normalized in the same \dphi{} range) was
subtracted.
 
\section{Results}

\subsection{Azimuthal di-hadron distributions}

\begin{figure*}[hbt!]

\includegraphics[width=\textwidth]{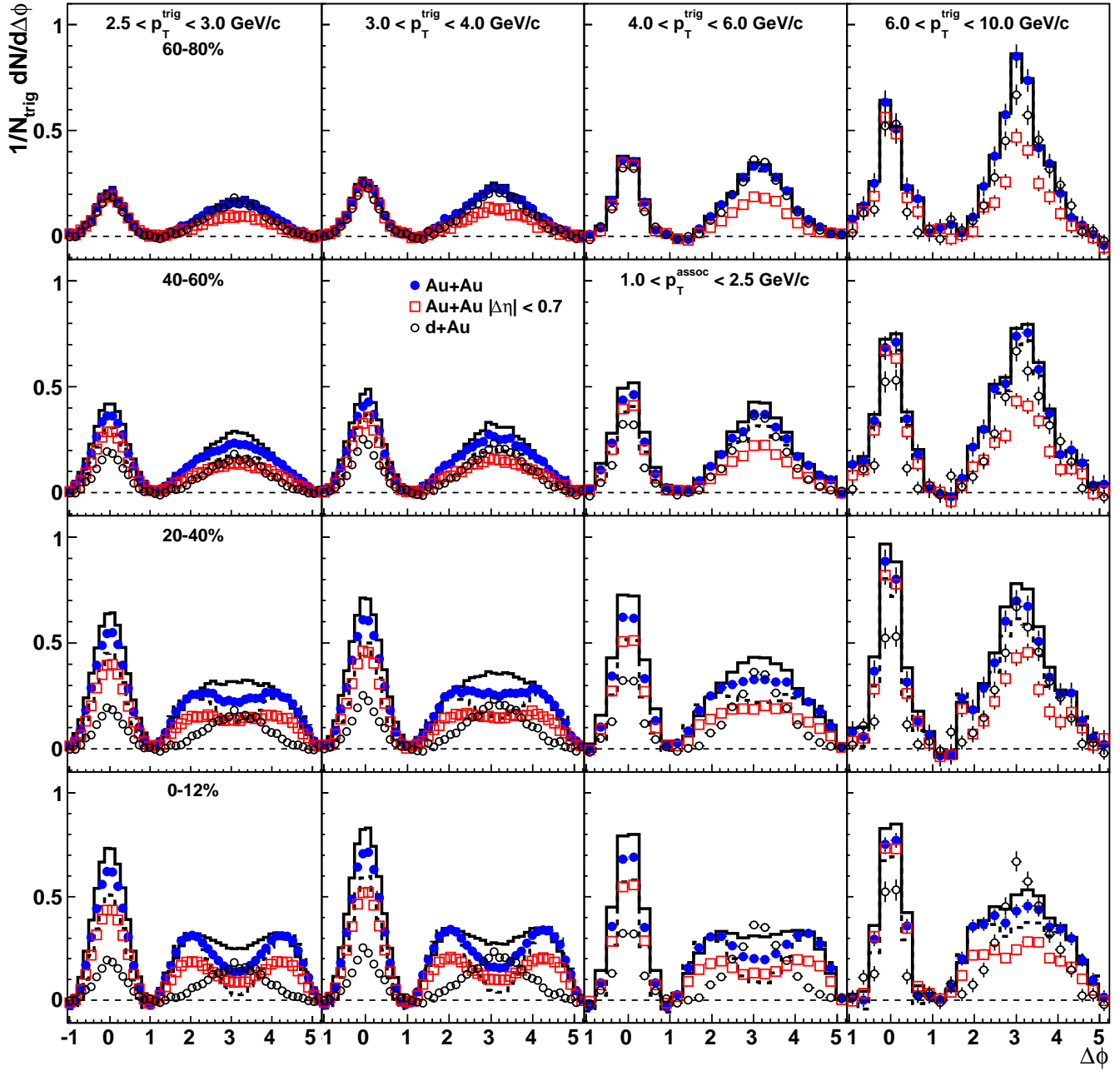}
\caption{(Color online) Background-subtracted azimuthal angle
difference distributions for associated particles with \pt{} between
1.0 and 2.5 GeV/$c$ and for different ranges of trigger particle
\pt{}, ranging from $2.5-3.0$ GeV/$c$ (left column) to $6-10$ GeV/$c$
(right column). Results are shown for Au+Au collisions (solid circles)
with different centrality (rows) and d+Au reference results (open
circles). The rapidity range is $|\eta|<1$ and as a result the
rapidity-difference $|\Delta\eta| < 2$. Open red squares show results
for a restricted acceptance of $|\Delta\eta|<0.7$, using tracks within
$|\eta|<1$. The solid and dashed histograms show the upper and lower
range of the systematic uncertainty due to the $v_2$ modulation of the
subtracted background.}
\label{fig:dphi_cent}
\end{figure*}

Fig.~\ref{fig:dphi_cent} shows the background subtracted associated
hadron \dphi{} distributions with $1.0\pTab 2.5$ \GeVc{} for 4
centrality selections, 60-80\%, 40-60\%, 20-40\% and 0-12\%, and 4
trigger selections, $ 2.5\pTtb3.0$ \GeVc, $ 3.0\pTtb4.0$ \GeVc, $
4.0\pTtb6.0$ \GeVc, and $6.0\pTtb10.0$ \GeVc.  Results are presented
for two different ranges in the pseudo-rapidity difference between the
trigger and associated particles $|\Delta\eta|$.  The shapes are very
similar for both $\Delta\eta$ selections in all panels (there is an
overall reduction in the away-side yields due to the smaller
acceptance for $|\Delta\eta|<0.7$). For reference,
the di-hadron distributions without background subtraction are shown
in the Appendix, where also the $v_2$ values and background
normalization values ($B$) used to subtract the background are
given. The systematic uncertainties on the $v_2$ values for the
background are shown by the bands around the data points. The d+Au
results (open circles) are also shown for reference.

In Fig.~\ref{fig:dphi_cent}, top row, one
observes that the jet-like correlations in peripheral (60-80\%
centrality) Au+Au collisions are very similar to the d+Au result,
indicating that such correlations in peripheral Au+Au collisions can
be described as a superposition of independent p+p collisions. The
near- and away-side yields of associated particles increase with
\pTtrig, as expected from parton fragmentation.

For more central events, a significant increase of both the near- and
the away-side yields is seen in Au+Au collisions relative to d+Au. The relative increase of the near-side yield is
larger for lower \pTtrig\ (top row) than for higher \pTtrig. For
peripheral events, the near-side results for $|\Delta\eta|<0.7$ do not differ
significantly from the full acceptance results, demonstrating that the
correlated yield is at relatively small \deta, as expected from jet
fragmentation. For more central collisions, on the other hand, a
significant fraction of the associated yield is at large $|\deta| >
0.7$ for the lower \pTtrig, indicating a significant
long-range correlation in \deta{}, possibly due to an interplay
between the soft bulk dynamics of longitudinal flow and jet-like
di-hadron structure \cite{Voloshin:2003ud}. It has
also been argued recently that this long-range correlation in \deta{}
could be caused by long-range structures in the medium, due to density
fluctuations in the medium \cite{Takahashi:2009na} or color flux tubes
\cite{Liao:2007mj,CGCRidge,Gavin:2008ev}.
The `ridge'-like correlation
structure in \deta{} is further explored in other STAR publications
\cite{Adams:2004pa,Adams:2006tj,Abelev:2009qa,Abelev:2009jv}.

It is interesting to note that the largest relative enhancement of the
near-side yield is observed for the lower \pTtrig, 2.5 -- 4.0
GeV/$c$. It has been suggested that particle production in this
momentum range has a large contribution from coalescence of quarks
from bulk partonic matter
\cite{Greco:2003xt,Hwa:2002zu,Fries:2003vb}. This production mechanism
would not lead to jet-like structures. Trigger hadrons formed by this
mechanism would increase the number of trigger hadrons, without
increasing the associated yield, leading to a reduced per-trigger
associated yield, in contrast to what is observed in
Fig. \ref{fig:dphi_cent}. The increased associated yield at
intermediate \pt{} indicates that if coalescence is a significant
source of hadron production at intermediate \pt, it has to generate
angular correlation structure, either through shower-thermal
coalescence \cite{Hwa:2004ng} or local fluctuations in the medium
density or temperature, e.g. due to heating of the medium by the
passage of a parton \cite{Chiu:2005ad}. So far, most calculations of
such effects are qualitative at best. Quantitative predictions for
these processes should be made and compared to the data. Measurements
with identified baryons and mesons as trigger and associated particles
\cite{Adare:2006nn,Adler:2004zd} can be used to explore the possible
contributions from coalescence.

The away-side yield of associated particles at low \pTtrig\ (top row
Fig. \ref{fig:dphi_cent}) evolves significantly in both shape and
yield with centrality: the shape becomes much broader than the d+Au
reference and the yield increases. For 20-40\% central collisions the
distribution becomes flat or slightly double-peaked, with a shallow
minimum at $\dphi=\pi$. In the most central collisions the
distribution is double-peaked for the lowest \pTtrig{}. With
increasing \pTtrig, the away-side shape becomes flatter. Overall,
there is a smooth evolution of the peak shape with centrality and
\pTtrig. The value of \pTtrig\ for which the away-side becomes flat or
double-peaked decreases with centrality. Note that the double-peak
shape is not seen in the raw signal in Fig. \ref{fig:dphi_bkg} and only
appears after the subtraction of the $v_2$-modulated background. In
that sense, the double-peak structure is generated by imposing a
separation between flow and non-flow in the analysis of azimuthal
correlations. This separation is not unambiguous and remains under
active investigation.

For the most central collisions, the broadening of the away-side
structure is so large that the near- and away-side peaks may overlap,
making it impossible to unambiguously distinguish the correlation
structure from the background without other inputs. For the present
analysis, we have chosen to use the same background normalization
procedure for all centrality bins and \pt{} bins, {\it i.e.} to
normalize the $v_2$-modulated background to the signal in the range
$0.8<|\dphi|<1.2$ and subtract it.

\begin{figure*}[hbt!]
\includegraphics[width=\textwidth]{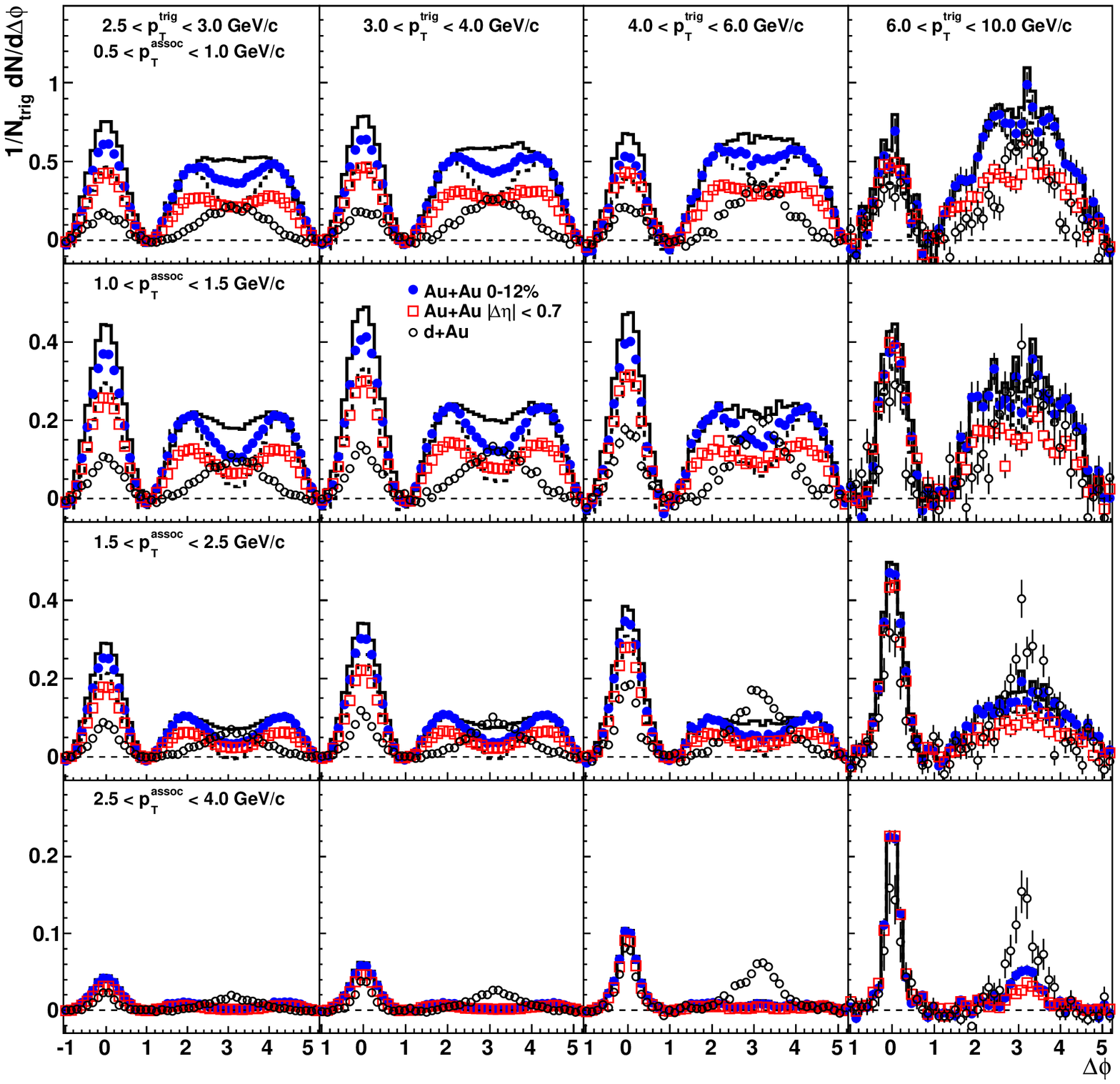}
\caption{(Color online) Background-subtracted azimuthal angle difference
distributions for different \pTtrig{} (columns) and \pTassoc{} (rows)
in 0-12\% central Au+Au collisions (solid circles) and d+Au reference
results (open circles). The rapidity range is
$|\eta|<1$ and as a result the rapidity-difference $|\Delta\eta| <
2$. Open red squares show results for a restricted acceptance of
$|\Delta\eta|<0.7$, using tracks with
$|\eta|<1$. The solid and dashed histograms show the upper and lower
range of the systematic uncertainty due to the $v_2$ modulation of the
subtracted background. }
\label{fig:dphi_pttrig}
\end{figure*}

In Fig.~\ref{fig:dphi_pttrig} we focus on central data where the
largest modifications of the correlation shapes and yields are found.
The figure shows the correlation shapes in the 0-12\% central event
sample for different selections of \pTassoc{} and \pTtrig{}. As in
Fig.~\ref{fig:dphi_cent}, results are given for the full
\deta-acceptance (solid circles) as well as a restricted range
$|\deta| < 0.7$ (squares), and for d+Au collisions (open
circles). The distributions before background
subtractions and the background normalization and $v_2$ values are
given in the Appendix.

On the near-side, we observe again a large increase of the yield in
central Au+Au collisions compared to d+Au collisions. The yield
depends on the \deta-selection used, indicating that there is
significant associated yield at $\deta > 0.7$. The relative size of
the enhancement depends on \pTassoc{} and \pTtrig. The measured jet-like yield in d+Au collisions increases faster with \pTtrig{} (going from left to right in Fig.~\ref{fig:dphi_pttrig}) than in Au+Au collisions, reducing the
relative size of the enhancement in Au+Au. The associated yield
decreases with increasing \pTassoc{} for both d+Au and Au+Au collisions, but
the decrease is stronger in Au+Au, so that the measured yields in Au+Au
approach the d+Au results at the highest \pTassoc. A summary of the
yields is presented in Fig.~\ref{fig:Yields} (Section \ref{sect:spectra}).

On the away-side, we observe a broadening and enhancement of the yield
in Au+Au compared to d+Au, except at $2.5 \pTab 4.0$ GeV/$c$ (bottom
row of Fig.~\ref{fig:dphi_pttrig}), where a broadening is seen, while
the yield is smaller than in d+Au. For $6 \pTtb 10$ GeV/$c$
(right-most column in Fig.~\ref{fig:dphi_pttrig}), a narrow peak
appears at large \pTassoc{} in Au+Au, similar to what is seen in d+Au collisions
and at higher \pt{} in Au+Au collisions \cite{Adams:2006yt}.

Although the shape of the away-side distribution changes with
\pTtrig{} and \pTassoc, there seems to be no gradual broadening as a
function of \pt: the rising flanks of the away-side distribution are
at similar \dphi{} in the entire range $0.5 <
\pTassoc < 2.5$ \GeVc{} and $2.5 < \pTtrig < 6$. In fact, it could be
argued that the away-side distribution is as broad
as possible; there is no \dphi{} region without correlation
signal.

\begin{figure*}[th!]
\includegraphics[width=0.8\textwidth]{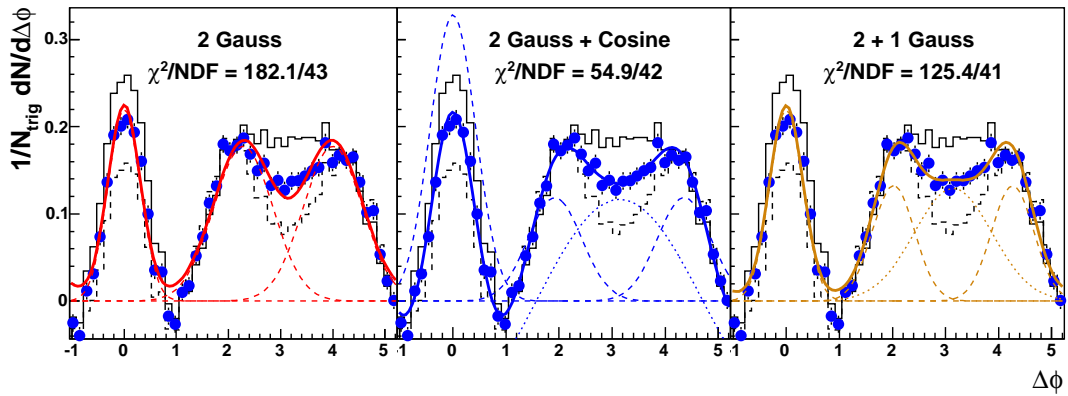} 
\caption{(Color online) Background-subtracted associated hadron
distribution in 0-12\% central Au+Au collisions with $0.8 \pTab 1.0 $
\GeVc{} and $4 \pTtb 6$ \GeVc{}. The colored curves indicate fits
with 3 different functional forms for the away-side shape: (left
panel) symmetric identical Gaussian distributions (2 Gauss), (middle
panel) adding a cos$\dphi$ background distribution (2 Gauss + Cosine),
and (right panel) adding a third Gaussian as the away-side jet-like
component (2+1 Gauss). The histograms indicate the
uncertainty on the background shape due to elliptic flow.}
\label{fig:FitDemo}
\end{figure*}

The broad away-side correlation structure in Au+Au collisions is a
truly remarkable observation. Although some
broadening of the away-side correlation in Au+Au collisions would be
expected due to increased acoplanarity ($k_{T}$) due to multiple
scattering of the parton in the medium, the structures seen in
Fig. \ref{fig:dphi_pttrig} are broader than would be expected from
such a mechanism \cite{Vitev:2005yg}. It has, however, been pointed
out that kinematic selection effects on in-medium gluon radiation may
lead to a non-trivial structure in the angular distributions
\cite{Polosa:2006hb}. It has also been argued that a fast parton may
generate sound waves in the bulk quark-gluon matter which would lead
to a Mach-cone shock wave
\cite{Stoecker:2004qu,CasalderreySolana:2004qm,Renk:2005si,Ruppert:2005uz}.
Evidence for a conical emission pattern has been
found in three-particle correlation measurements
\cite{Abelev:2008nd}. The broad structure seen in the di-hadron
distribution could then be the projection of the conical pattern on
$\dphi$. Another mechanism that may produce conical emission from a
fast parton is QCD Cherenkov radiation
\cite{Majumder:2005sw,Koch:2005sx}. There are two
other calculations that show a broad, double-peaked away-side
structure without implementing a specific mechanism for conical
emission: one is a 3D hydrodynamical calculation which includes local
density fluctuations in the initial state \cite{Takahashi:2009na} and
the other is the AMPT model \cite{PhysRevC.76.014904}. It is worth
noting that in the 3D hydrodynamical model, there is also no explicit
introduction of hard partons or jets; the correlation arises purely
from the medium.  All these models should be confronted with the data
presented in Figs~\ref{fig:dphi_cent} and \ref{fig:dphi_pttrig}
as well as the three-particle correlation data in
\cite{Abelev:2008nd}.

In addition to the change of the correlation shapes, a significant
increase of the yields is seen in Au+Au collisions relative to d+Au collisions, for most of the \pt-selections in
Fig. \ref{fig:dphi_pttrig}, both on the near- and the away-side. The
yield increase implies that trigger hadrons in Au+Au collisions are
accompanied by a larger energy flow than trigger hadrons with the same
transverse momentum in elementary collisions. This would be compatible
with a scenario where the leading hadron is softened due to energy
loss so that trigger hadrons in Au+Au collisions select a larger
initial parton energy than in d+Au or p+p collisions.

\subsection{Away-side shapes}

To further characterize the broad shape of the away-side associated
hadron distributions in Figs~\ref{fig:dphi_cent}
and~\ref{fig:dphi_pttrig}, the data were fitted with different
parameterizations. Three different functional forms were used, all
of which are based on the assumption that there is
significant yield in a cone in (\deta,\dphi) around the away-side
parton. The projection of this cone on $\dphi$ would give rise to two
peaks symmetric around
$\dphi=\pi$. Fig. \ref{fig:FitDemo} shows the
associated hadron distribution after background subtraction for $4
\pTtb 6$ \GeVc{} and $0.8 \pTab 1.0$ \GeVc{} fitted with three
different functional forms that include two Gaussian distributions
at $\dphi = \pi \pm \Delta$ on the away-side. The left panel of
Fig. \ref{fig:FitDemo} shows the simplest ansatz, using just two
Gaussian peaks on the away side and a single Gaussian peak on the near
side. To account for correlations induced by momentum
conservation or remnant jet structure, we add a $\cos(\dphi)$
distribution (middle panel of Fig.~\ref{fig:FitDemo}) or a third
Gaussian peak (with a different width and amplitude)
at $\dphi=\pi$ (right panel
Fig. \ref{fig:FitDemo}). The three parameterizations
are referred to as 2 Gauss, 2 Gauss + Cosine and 2+1 Gauss,
respectively. The best fit is obtained when using the 2 Gauss + Cosine
function, as can be seen from the $\chi^2$ values given in the
figure.

\begin{figure}
\includegraphics[width=0.49\textwidth]{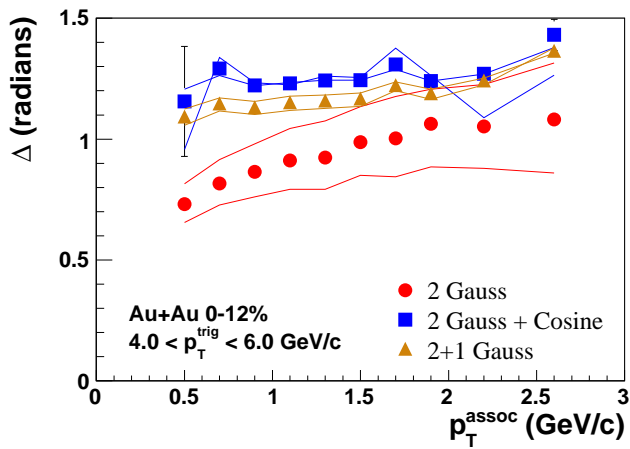}
\caption{(Color online) The angle $\Delta$ between the away-side peaks and
$\dphi=\pi$ for $4<\pTtrig<6$ \GeVc{} in 0-12\% central Au+Au
collisions as a function of \pTassoc. Three different parametrisations
of the away-side peak shape were used (see text). The lines show the systematic
uncertainty from $v_2$ variation while the errors are statistical
errors from the fit.}
\label{fig:PeakLocation}
\end{figure}

\begin{figure*}[th!]
\begin{minipage}{0.49\textwidth}
\centering
\includegraphics[width=\textwidth]{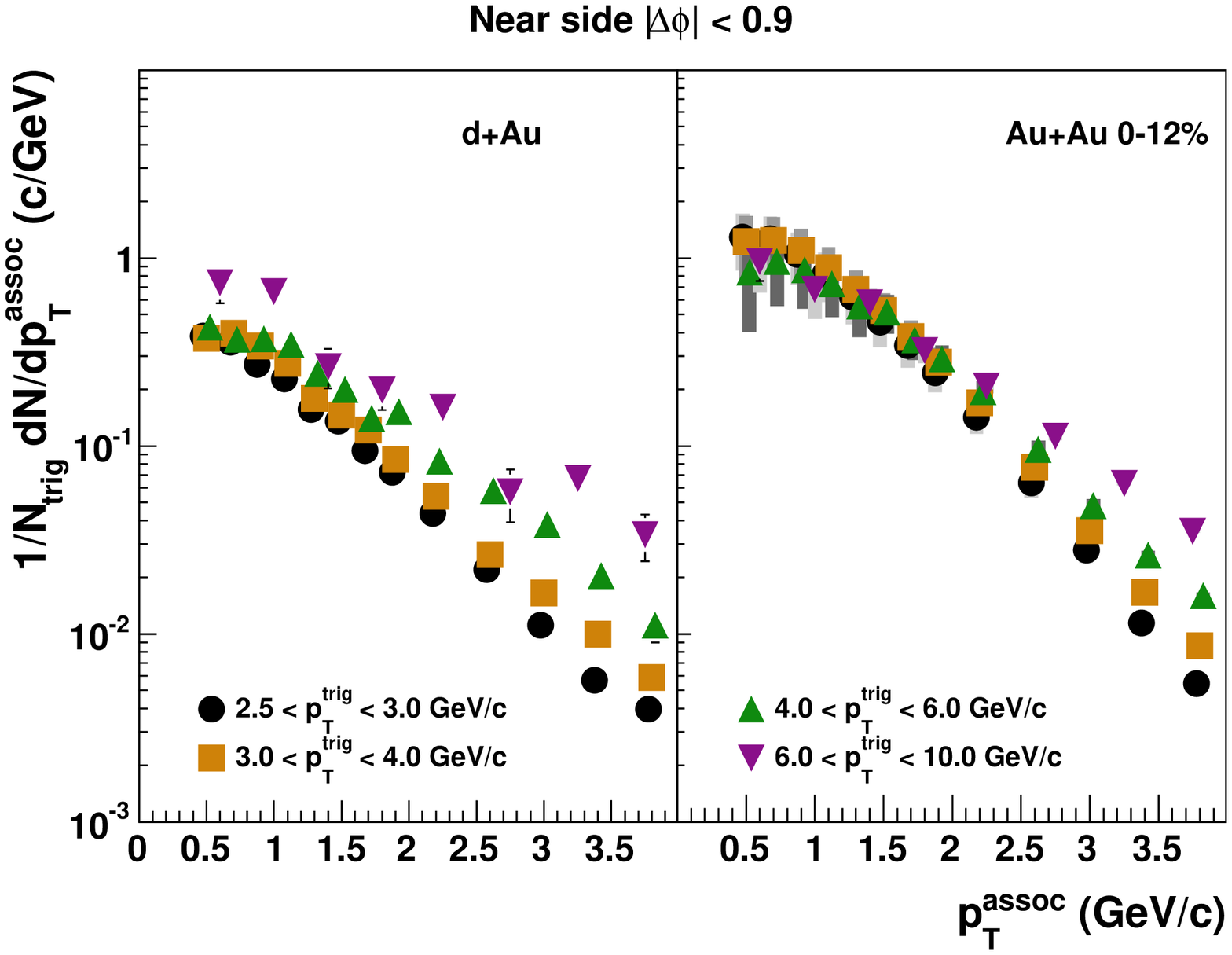}
\includegraphics[width=0.9\textwidth]{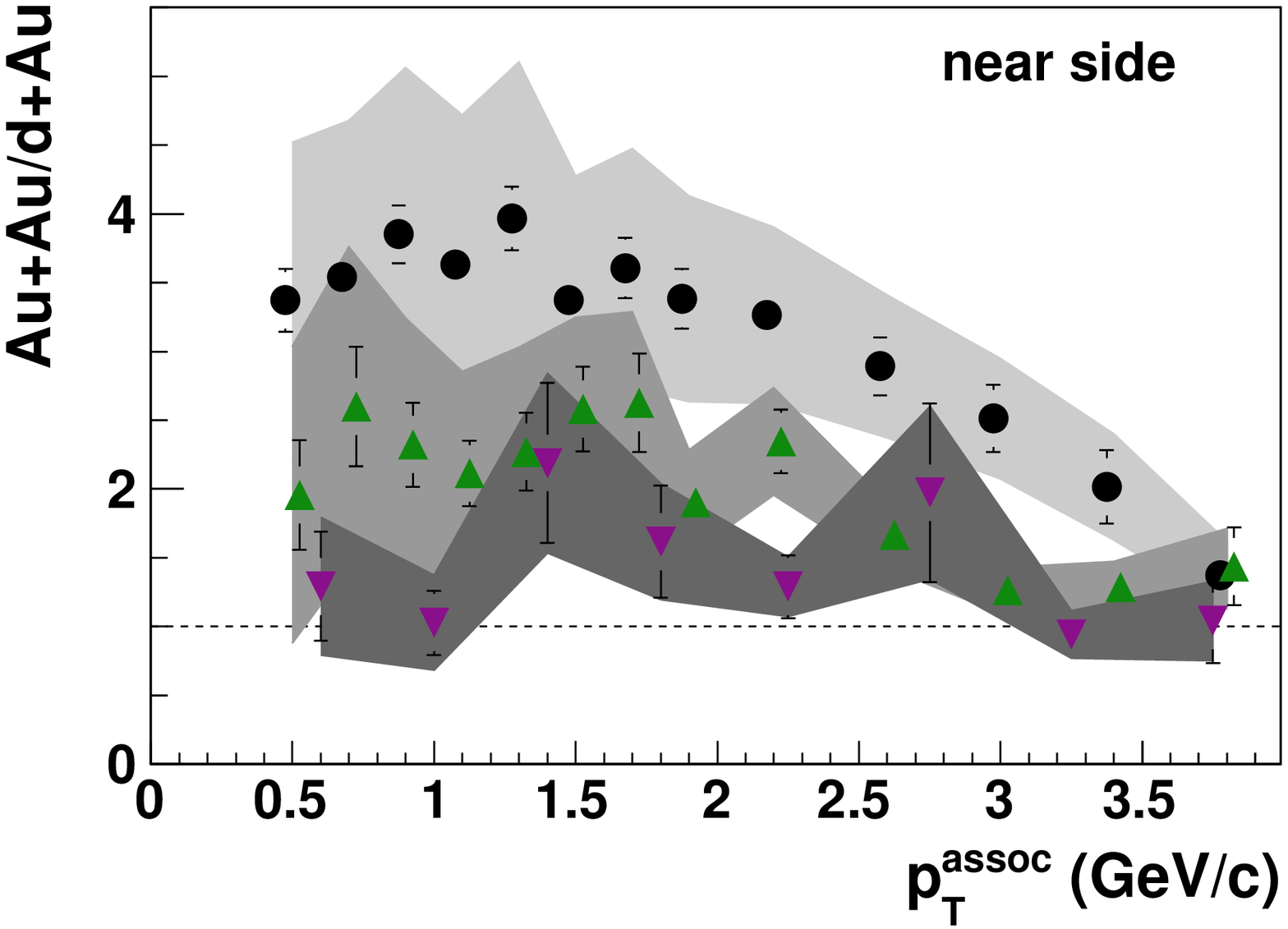}
\end{minipage}\hfill%
\begin{minipage}{0.49\textwidth}
\centering
\includegraphics[width=\textwidth]{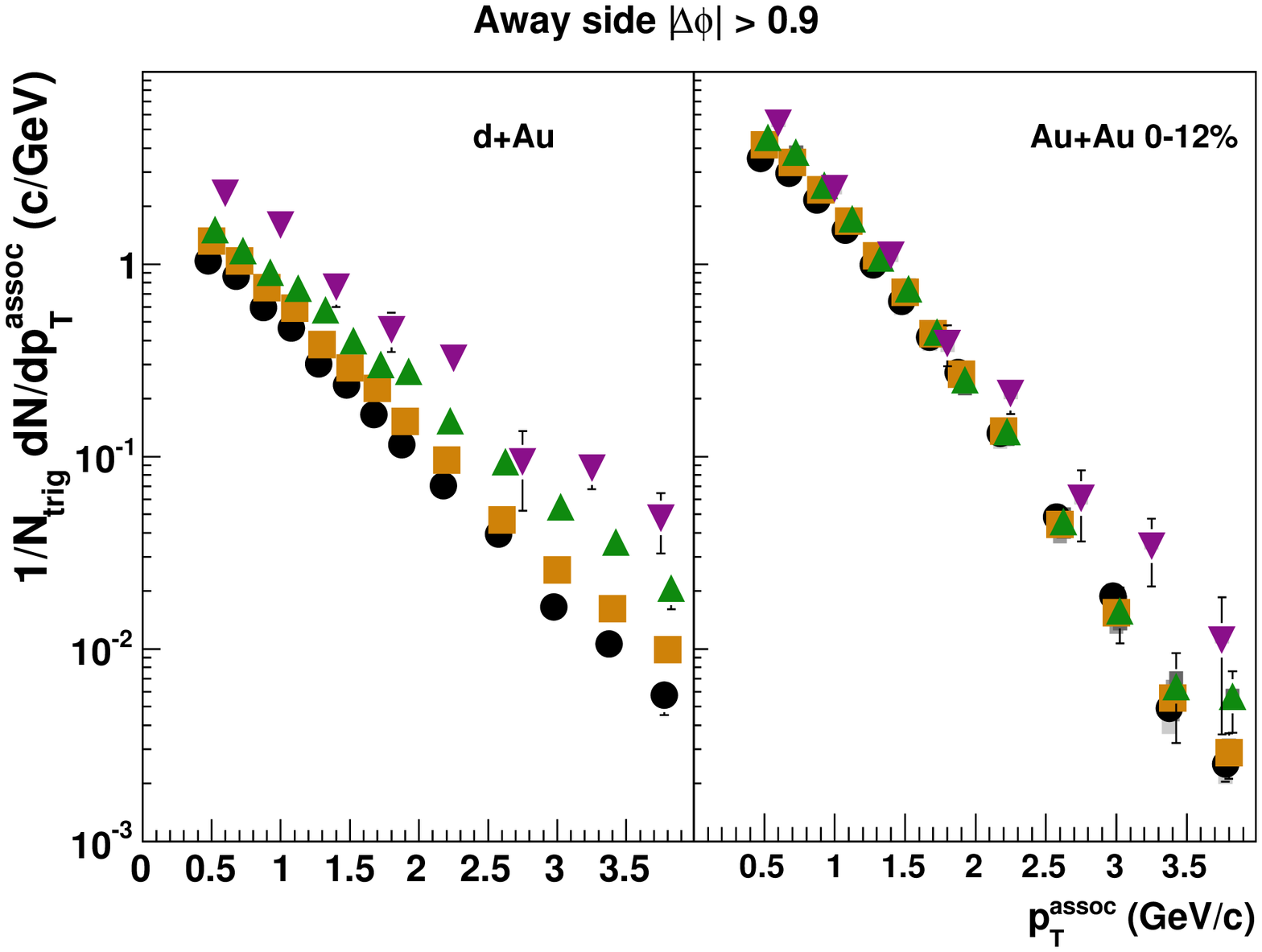}
\includegraphics[width=0.9\textwidth]{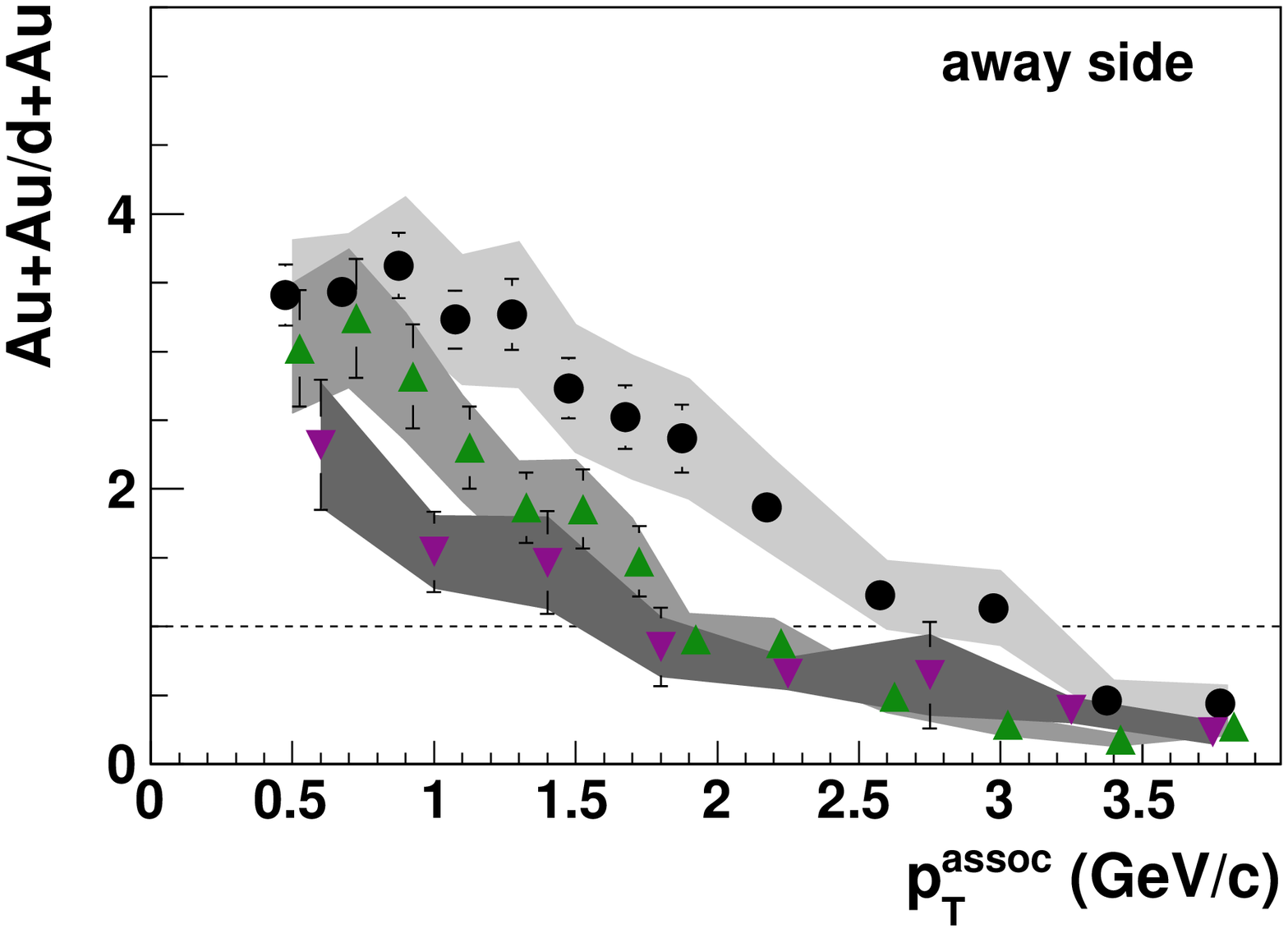}
\end{minipage}
\caption{(Color online) \label{fig:Yields}Near-side ($|\dphi| < 0.9$,
  left panels) and away-side ($|\dphi| > 0.9$, right panels)
  associated yield per trigger particle for various \pTtrig\
  selections as a function of \pTassoc. Results are shown for 0-12\%
  central Au+Au collisions and for d+Au collisions. The bottom panels
  show the ratios of the per-trigger associated yields in central
  Au+Au and d+Au collisions. The error bars on the points indicate the
  statistical uncertainty, including the statistical uncertainty on
  the subtracted background level. The gray bands indicate the
  uncertainty from the elliptic flow modulation of the background.}
\end{figure*}

Figure~\ref{fig:PeakLocation} shows $\Delta$, the angle (in radians)
between the Gaussian peaks and $\dphi=\pi$, as a function of
\pTassoc, for the three different parametrizations of the away-side
shape for $4 < \pTtrig < 6$ \GeVc{} 
in the 0-12\% most central
collisions. Similar results were obtained for $3 < \pTtrig{} < 4$ \GeVc{} (not shown).

The peak positions $\Delta$ in Fig.~\ref{fig:PeakLocation} show a slow
increase with \pTassoc{} for the fits with the Symmetric Gaussian
form. This functional form alone however, does not provide a good
description of the away-side shape for larger \pTtrig{} and
\pTassoc{}. When an away-side contribution at $\dphi=\pi$ is included
(2+1 Gauss and and 2 Gauss+Cosine shapes), the peak position $\Delta$
is close to 1.2 and approximately independent of \pTassoc. This
observation is qualitatively consistent with predictions for a Mach Cone
developing in the hot and dense medium of the early stage of the
collision ~\cite{CasalderreySolana:2004qm} and with existing results
on three-particle azimuthal correlations \cite{Abelev:2008nd}.

\subsection{Associated particle spectra}
\label{sect:spectra}

Figure~\ref{fig:Yields} shows the integrated yield in the near-side
peak ($|\dphi| < 0.9$) and away-side ($|\dphi| > 0.9$) as a function
of \pTassoc{} in central Au+Au collisions and d+Au collisions for four
\pTtrig{} intervals.  
The lower panels of the
figures show the ratio of the associated yields in central Au+Au
collisions and d+Au collisions.

For d+Au collisions, the associated yield clearly increases with increasing
\pTtrig{} and the slope decreases with \pTtrig. These trends are
expected from parton fragmentation, where the larger \pTtrig{} selects
larger underlying parton
energies, thus increasing the multiplicity in the jet and leading to a
harder fragmentation.

The lower panels of Fig. \ref{fig:Yields} show that the ratio of the
yields in Au+Au and d+Au is decreasing with \pTassoc{}, indicating
that the fragmentation is softened due to in-medium energy loss. A
softer fragmentation also implies that a trigger particle of given
momentum selects different parton energies in Au+Au collisions than in
d+Au collisions, which could explain some of the enhancement of
associated yield at lower \pTassoc{} in Au+Au collisions. However, it
should be noted that a large part of the increased yield at lower
\pTassoc{} is at large \deta{}, associated with the ridge effect, and
is therefore not necessarily from jet fragments \cite{Abelev:2009qa}.
It is also
possible that other sources of particle production, such as parton
coalescence and resonance decays, contribute at lower \pTtrig{} and
may lead to different behaviour in d+Au and Au+Au.

\subsection{Azimuthal angle dependent mean-\pt}

\begin{figure}[hbt!]
\includegraphics[width=\columnwidth]{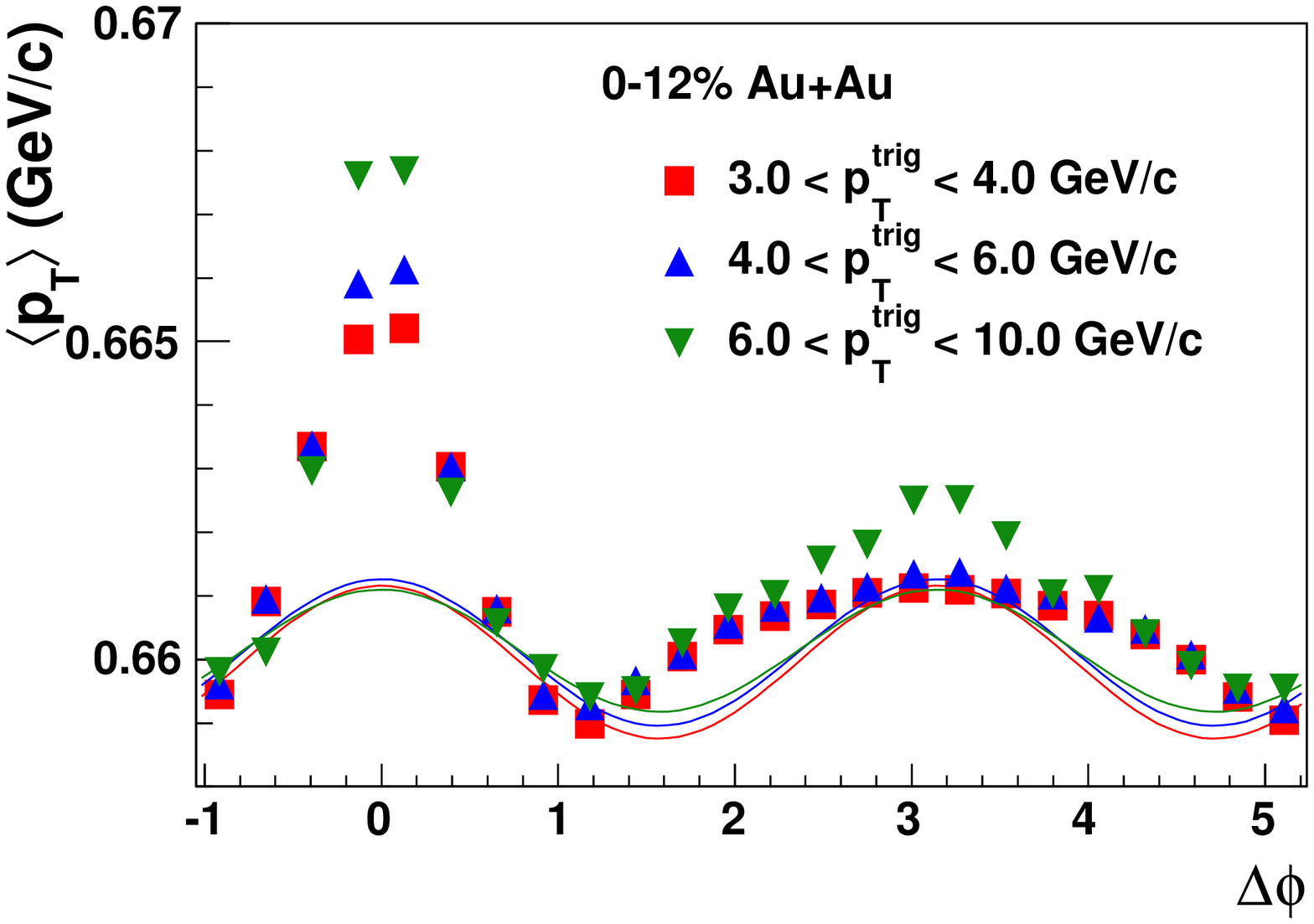}
\caption {(Color online) Inclusive, azimuthal $\mpt$ of associated hadrons between
  0.25 and 4.0 \GeVc{} in all events containing various classes of
  trigger particles. The lines show the $v_2$-modulated background for
  the different trigger ranges, with the colors corresponding to the
  data points.}
\label{fig:InclusiveMean}
\end{figure}

To further characterize the \pt-dependence of associated particle
production, we perform an analysis of the inclusive mean-\pt{}, \mpt,
of associated particles as a function of \dphi{}. The azimuthal
distribution $\mpt(\dphi)$ is calculated by
taking the ratio of the \pt-sum distribution $P_{T}(\dphi)$ and
the number distribution $N(\dphi)$
\begin{equation}
\mpt(\dphi)=P_{T}(\dphi) / N(\dphi).
\label{eq:mpt}
\end{equation}
The number distribution $N(\dphi)= 1/N_{trig} dN/d\Delta\phi$, as
shown in Fig. \ref{fig:dphi_bkg}, while the \pt-sum distribution
$P_{T}(\dphi)$ is formed using the same procedure, but addding the
(scalar) transverse momenta as weights in the azimuthal distribution.

To illustrate this method, Fig.~\ref{fig:InclusiveMean} shows the
inclusive distribution $\mpt(\dphi)$ for 0.25 \pTab 4.0~\GeVc{} and
three different \pTtrig{} selections for 0-12\% central collisions. On
the near-side, a clear increase of $\mpt$ with \pTtrig{} is visible,
while the away-side $\mpt$ distributions show a smaller
dependence on \pTtrig{}. The lines in Fig.~\ref{fig:InclusiveMean} show the
background. The elliptic flow of the background is calculated as a
weighted average of $v_2^{trig}\,v_2^{assoc}$. The difference between
the \pt-weighted average $\langle v_2 \rangle_{p_{T}}$, which is used to subtract the
background in the \pt-weighted distribution, and the number-weighted
$\langle v_2 \rangle_N$, gives rise to the flow modulation of the background shown in the
figure.

\begin{figure*}[hbt!]
\includegraphics[width=\textwidth]{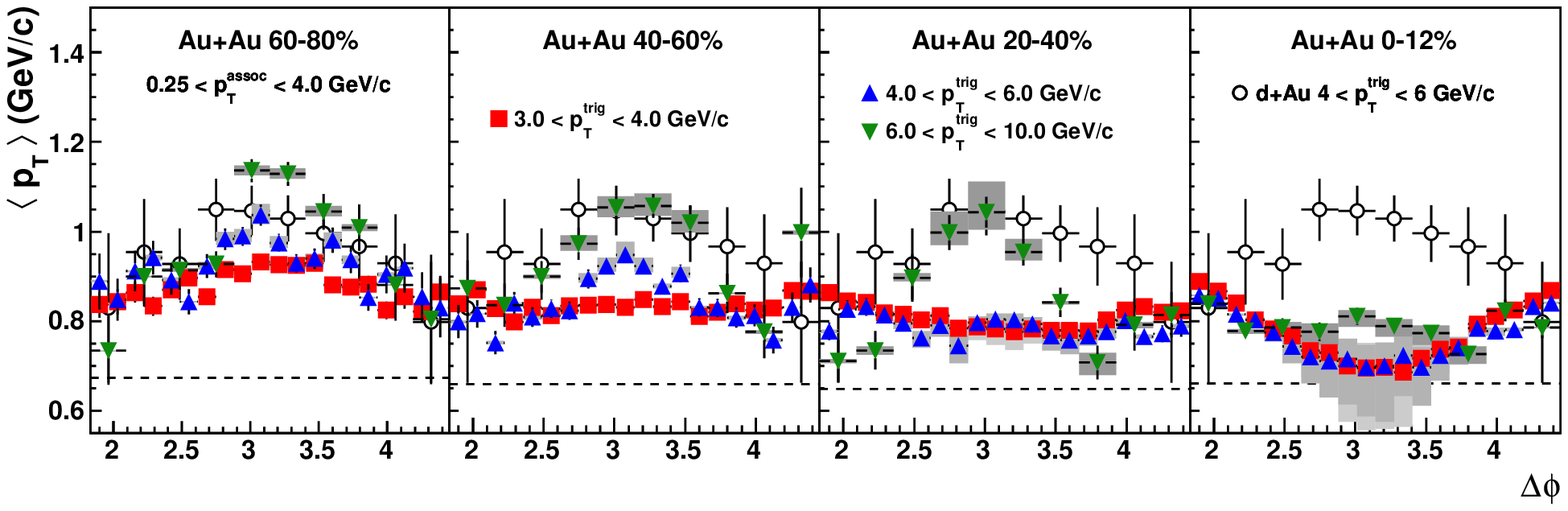}
\caption{(Color online) Mean transverse momentum $\mpt$ of associated particles with
 0.25 \pTab{} 4.0 \GeVc, for four different
 centrality selections. The shaded bands show the systematic uncertainty due
 to elliptic flow of the uncorrelated background. The dashed lines indicate the inclusive $\mpt$ in the same \pt{} range in events with a trigger particle.}
\label{fig:CorrMean}
\end{figure*}

To calculate the \mpt{} of associated hadrons, the uncorrelated
background is subtracted from both the \pt-weighted and
number-weighted distributions before taking the
ratio:
\begin{equation} 
\mpt(\dphi)=\frac{P_{T}(\dphi)-B_{p_T} ( 1 + 2 \langle v_2 \rangle_{p_{T}}\cos(\dphi))}{N(\dphi) - B_{N} ( 1 + 2 \langle v_2 \rangle_N \cos(\dphi))},
\end{equation}
where $N(\dphi)$ and $P_{T}(\dphi)$ are the same
number-weighted and sum-\pt{} distributions used in Eq. \ref{eq:mpt}, the
average $\langle v_{2} \rangle$ are defined above and $B_{p_{T}}$ and
$B_{N}$ are background normalizations which are determined using the
ZYAM method separately for the number and sum-\pt{} distributions.

Figure~\ref{fig:CorrMean} shows the resulting $\mpt$ of associated
hadrons as a function of \dphi{} in the away-side
region for different centrality selections. In the peripheral bins a
peaked structure in $\mpt$ is found, similar to the results in
d+Au collisions (open circles).  With increasing centrality, the
\mpt{} around $\dphi = \pi$ becomes lower. For the most central bin,
the results show a minimum at $\dphi = \pi$ for the two softer trigger
selections. For the highest trigger selection $6 \pTtb 10$
GeV/$c$, a similar shape is seen, but there may be a slight enhancement
at $\dphi=\pi$ even in the most central collisions.

\begin{figure}[hbt!]
\includegraphics[width=0.45\textwidth]{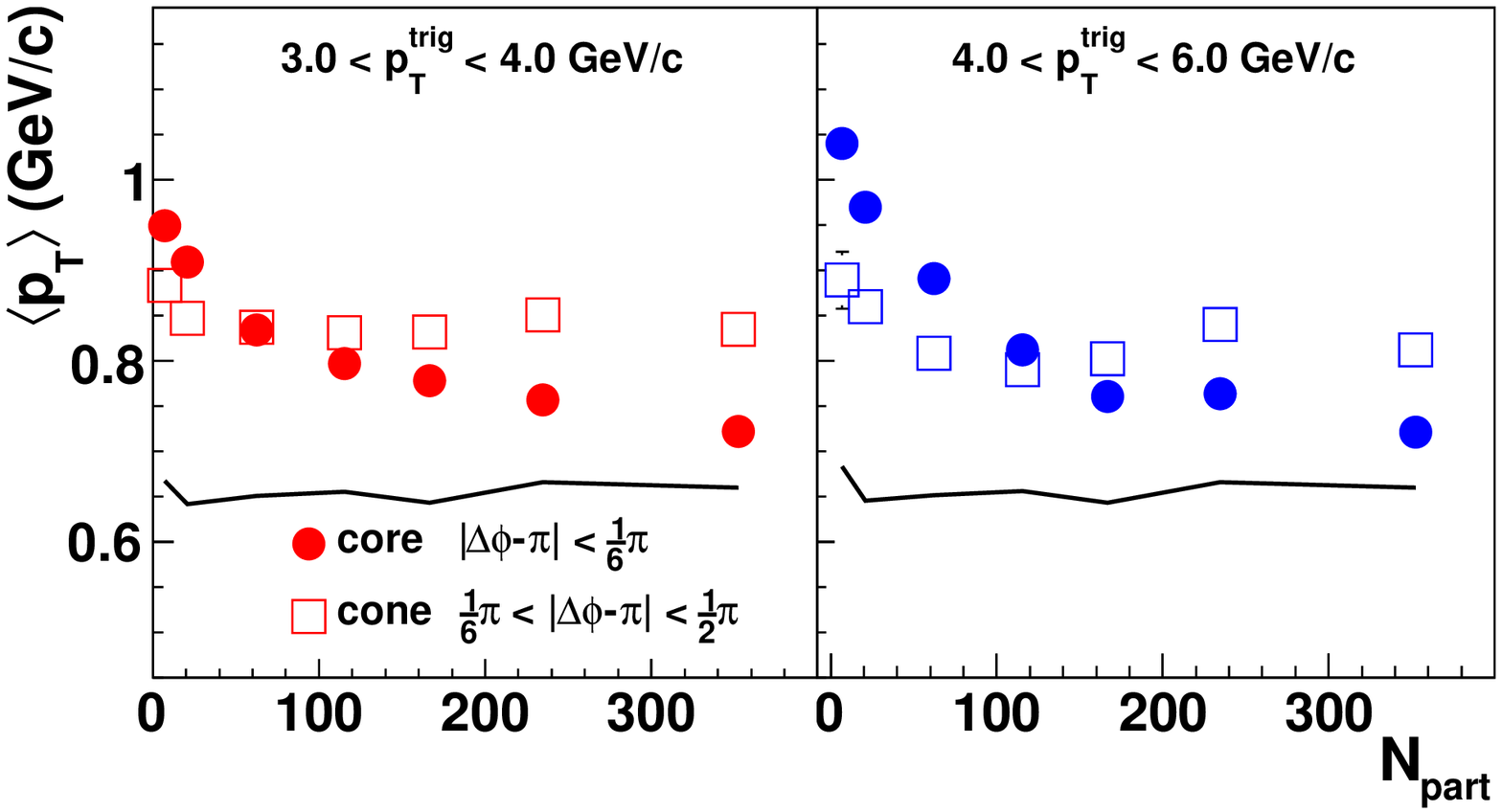}
\caption{(Color online) Mean transverse momentum $\mpt (\dphi)$ of associated hadrons in the range
$0.25 < \pTassoc < 4.0 $ \GeVc{} in the core ($|\dphi-\pi| < \frac{\pi}{6}$)
and cone ($\frac{\pi}{6} < |\dphi-\pi| < \frac{\pi}{2}$) azimuthal
regions for $3 < \pTtrig < 4$ \GeVc{} and $4 < \pTtrig < 6$ \GeVc{} as
a function of number of participants. The lines indicate the inclusive
\mpt{} in the same \pt{} range for events with a trigger hadron. }
\label{fig:NpartDep}
\end{figure}

We further study the difference between $\mpt$ in the range
$|\dphi-\pi| < \frac{\pi}{6}$ (referred to as ``core" in the
following) and at $\frac{\pi}{6} < |\dphi-\pi| < \frac{\pi}{2}$
(referred to as ``cone" in the following). Figure \ref{fig:NpartDep}
shows $\mpt$ in these two angular ranges as a function of the
collision centrality for two different trigger $p_T$ ranges.  The
$\mpt$ decreases with centrality approaching the inclusive $\mpt$, a
feature already reported in~\cite{Adams:2005ph} for associated hadrons
in the entire away-side region ($|\dphi-\pi|<2.14$). This reduction of
\mpt{} is likely due to interactions with the medium. The fact that
\mpt{} approaches the \mpt{} for inclusive particle production
in events with a trigger hadron (solid lines in
Fig.~\ref{fig:NpartDep}) may indicate that the associated particles at
low \pt{} approach thermalisation with the medium. It is also clear
that the \mpt{} of the core decreases more rapidly than that of the
cone hadrons, which suggests that there is significant transport of
associated hadrons away from $\dphi=\pi$ due to
jet-medium interactions.

\section{Discussion and conclusions}
In this paper, a comprehensive study of centrality and \pt-dependence
of azimuthal di-hadron correlations in Au+Au events is presented. We
observe several striking modifications of the correlation structure in
Au+Au compared to a d+Au reference. Associated yields on the near- and
away-side are enhanced at lower \pt. On the near-side, the increase in
yield is partly located at large pseudo-rapidity difference \deta{}
(see \cite{Abelev:2009qa} for a more detailed study) and the yields
approach the measurement in d+Au collisions at the highest \pTtrig. On
the away-side, the associated hadron distribution is significantly
broadened; in fact, it is broad enough that it is impossible to
unambiguously separate jet-like yields from the underlying event. At
higher \pt{}, $\pTassoc{}> 2$ GeV/$c$ and $\pTtrig >
6$ GeV/$c$, the away-side shape is narrow, like in d+Au events. A
large enhancement of the away-side yield at low \pt{} is found, while
at higher \pt{} a suppression is seen with respect to d+Au collisions.

These results are qualitatively consistent with a softening of
jet fragmentation by in-medium energy loss, leading to an increase of
the underlying parton energy selected by a trigger particle at given
\pTtrig{}. Some of the changes in the correlation shapes could then be due
to fragmentation of radiated gluons.

The strong broadening of the away-side shapes, however, does not seem
to fit 
naturally in a description of particle production from medium
modified jet fragmentation. Several alternative mechanisms have been
proposed that could give rise to these structures. These can be divided
into two categories: collective and radiative phenomena.

Radiative
treatments~\cite{Vitev:2005yg,Polosa:2006hb,Dremin:1979yg,Majumder:2005sw,Koch:2005sx}
focus on the angular distribution of gluons radiated by the
parton propagating through the medium. A large opening angle between
the parent parton and radiated gluons is expected when kinematic
constraints are imposed. Two simplified calculations of this effect
have been published in the literature. One of these calculations 
gives results that are qualitatively consistent with the 
data~\cite{Polosa:2006hb}, while the other calculation
\cite{Vitev:2005yg} shows a much smaller effect. Neither of the two
calculations includes full integration over the initial state kinematics
and the medium density development. Another radiative scenario
involves Cherenkov radiation of
gluons~\cite{Dremin:1979yg,Majumder:2005sw,Koch:2005sx}, which would
give rise to conical distributions.

Both for large-angle medium-induced gluon radiation and for gluon
Cherenkov radiation, the expectation is that the away-side shape
becomes narrower with increasing \pTassoc{}
\cite{Polosa:2006hb,Koch:2005sx}. This trend is not observed in the
correlation data: using a few different functional forms for the
away-side distributions, we found that peak-separation $\Delta$ is
approximately independent of \pTassoc. 

Alternatively, one could imagine that the passage of high-\pt{}
partons excites sound waves in the medium. It has been suggested that
this may lead to Mach shock waves
\cite{Stoecker:2004qu,CasalderreySolana:2004qm,Renk:2005si,Ruppert:2005uz}.
Qualitatively, the observed constant separation between the away-side
peak and the constant conical emission angle from
three-particle correlations \cite{Abelev:2008nd} are consistent with
this explanation. The transition from a broad away-side structure at
low \pt{} to a narrow structure at higher \pt{} would then signal the
change from away-side structures dominated by bulk particle production
from the medium to a situation where jet-fragments dominate.

A recent 3D hydrodynamical calculation which includes
local density fluctuations in the initial state also shows a broad
away-side structure that may be consistent with the experimental di-hadron correlation data
\cite{Takahashi:2009na}. In this model, there is no explicit
introduction of hard partons or jets; the correlation arises purely
from the medium. At the moment, it is not clear whether this model 
will also generate the conical emission signal seen in three-particle correlation
data \cite{Abelev:2008nd}. A study of three-particle correlations in this model is ongoing \cite{Andrade:2009em}. 

In general, a number of different mechanisms, including fragmentation,
radiative energy loss, bulk response and hadron formation by
coalescence of constituent quarks, may contribute to the observed
di-hadron correlation structures. Quantitative modeling of the
different processes, including the azimuthal
correlation of the trigger and associated hadrons with the reaction
plane, is needed to further disentangle the observed signals
and the background.

Experimentally, more insight in the underlying production processes
will be gained from di-hadron measurements with identified
particles and with respect to the reaction plane. In addition, $\gamma$-jet measurements, and
measurements with reconstructed jets are being pursued, which provide
better control over the initial state kinematics.

\section{Acknowledgements}
We thank the RHIC Operations Group and RCF at BNL, the NERSC Center at
LBNL and the Open Science Grid consortium for providing resources and
support. This work was supported in part by the Offices of NP and HEP
within the U.S. DOE Office of Science, the U.S. NSF, the Sloan
Foundation, the DFG cluster of excellence `Origin and Structure of the
Universe' of Germany, CNRS/IN2P3, STFC and EPSRC of the United
Kingdom, FAPESP CNPq of Brazil, Ministry of Ed. and Sci. of the
Russian Federation, NNSFC, CAS, MoST, and MoE of China, GA and MSMT of
the Czech Republic, FOM and NWO of the Netherlands, DAE, DST, and CSIR
of India, Polish Ministry of Sci. and Higher Ed., Korea Research
Foundation, Ministry of Sci., Ed. and Sports of the Rep. Of Croatia,
Russian Ministry of Sci. and Tech, and RosAtom of Russia.

\bibliographystyle{apsrev4-1}
\bibliography{star_dihadr_PRC07}

\appendix*
\section{Background shapes}
In this Appendix, we show the associated hadron azimuthal
distributions before subtracting the flow background.

\begin{figure*}[hbt!]
\includegraphics[width=\textwidth]{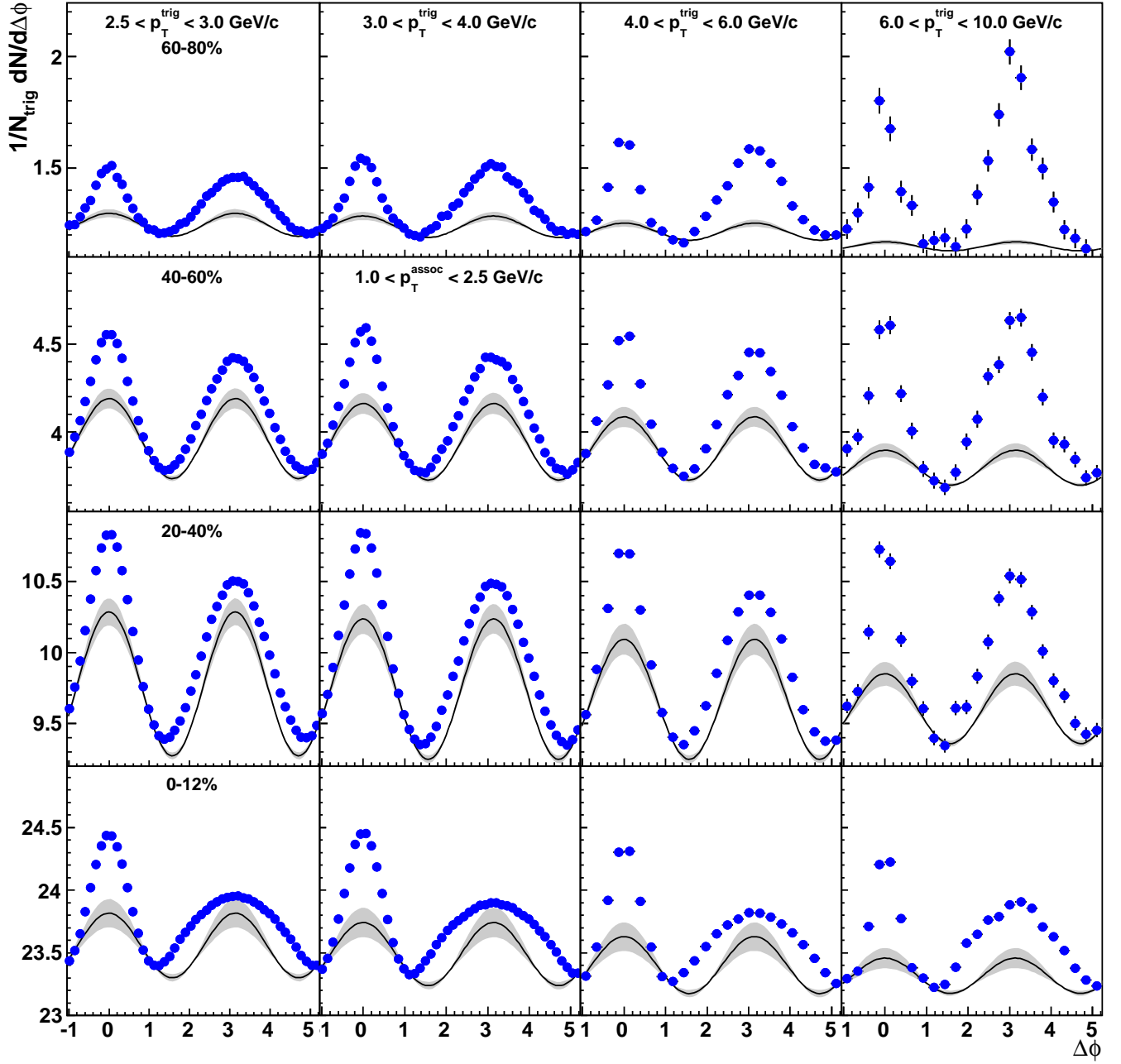}
\caption{(Color online) Azimuthal angle difference distributions for associated
particles with \pt{} between 1.0 and 2.5 GeV/$c$ and for different
ranges of trigger particle \pt{}, ranging from $2.5-3.0$ GeV/$c$ (left
column) to $6-10$ GeV/$c$ (right column). Results are shown for Au+Au
collisions with different centrality (rows). The line and the grey
band show the elliptic flow modulated background that was subtracted
to obtain Fig. \ref{fig:dphi_cent}.
\label{fig:dphi_cent_bkg}}
\end{figure*}

Figure \ref{fig:dphi_cent_bkg} shows the distributions for associated
particles between 1.0 and 2.5 GeV/$c$ and for different ranges of
trigger particle \pt{}, ranging from $2.5-3.0$ GeV/$c$ (left column)
to $6-10$ GeV/$c$ (right column) and different centralities
(rows). Note the large increase of the combinatorial background with
centrality. A large background modulation due to elliptic flow is
expected.

\begin{figure*}[hbt!]
\includegraphics[width=\textwidth]{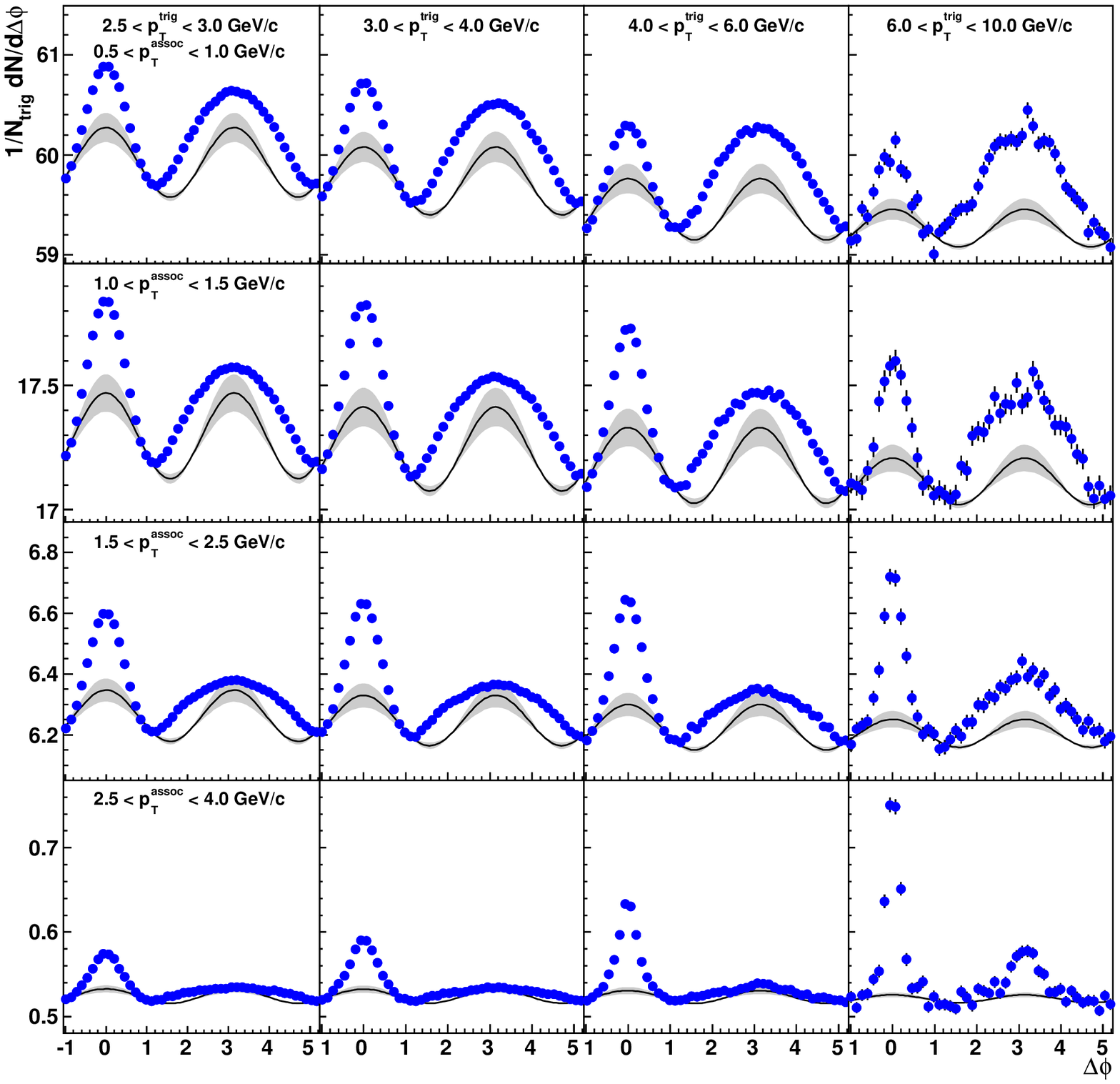}
\caption{(Color online) Azimuthal angle difference
distributions for different \pTtrig{} (columns) and \pTassoc{} (rows)
in 0-12\% central Au+Au collisions. The line and the grey band show the elliptic flow modulated background that was subtracted to obtain Fig. \ref{fig:dphi_pttrig}. }
\label{fig:dphi_pttrig_bkg}
\end{figure*}

Figure \ref{fig:dphi_pttrig_bkg} shows the distributions of associated
charged particles with various \pTassoc{} and \pTtrig{} selections for
0-12\% central Au+Au collisions. For $\pTassoc > 1.0$ GeV/$c$ (lower
three rows), the value of $1/N_{trig} dN/d\Delta\phi$ at the minimum
depends mostly on \pTassoc, as expected for uncorrelated background.
In the upper row however, with $0.5 < \pTassoc < 1.0$ GeV/$c$, we
observe a significant dependence of $1/N_{trig} dN/d\Delta\phi$ at
the minimum on \pTtrig. The value at the minimum increases for lower
\pTtrig, which is qualitatively consistent with 
a centrality bias combined with the fact that the probability to find 
more than one trigger particle per event is sizable for the lower \pTtrig{} 
selections and decreases with increasing \pTtrig{}.

\begin{table*}[htb] 
\begin{minipage}{\columnwidth}
\begin{tabular}{ccc}
\pTtrig\ & $B_{|\Delta\eta|<2.0}$ & $\langle v^{assoc}_2\rangle\langle v^{trig}_2\rangle$
	  \\ 
( GeV/$c$ ) & & ( $10^{-3}$ ) \\
 \hline
\multicolumn{3}{c}{ 60-80\% } \\ 
2.5 - 3.0  & 1.220 $\pm$ 0.002 & 20.6 $\pm$ 4.6 \\
3.0 - 4.0  & 1.214 $\pm$ 0.003 & 19.4 $\pm$ 4.6 \\
4.0 - 6.0  & 1.197 $\pm$ 0.008 & 15.7 $\pm$ 4.1 \\
6.0 - 10.0 & 1.139 $\pm$ 0.026 & 8.67 $\pm$ 2.7 \\
\multicolumn{3}{c}{ 40-60\% } \\ 
2.5 - 3.0  & 3.846 $\pm$ 0.002 & 28.6 $\pm$ 4.4 \\
3.0 - 4.0  & 3.835 $\pm$ 0.003 & 27.5 $\pm$ 4.6 \\
4.0 - 6.0  & 3.820 $\pm$ 0.008 & 22.9 $\pm$ 4.5 \\
6.0 - 10.0 & 3.76 $\pm$  0.03  & 13.0 $\pm$ 3.2 \\
\multicolumn{3}{c}{ 20-40\% } \\ 
2.5 - 3.0  & 9.522 $\pm$ 0.002 & 25.9 $\pm$ 3.0 \\
3.0 - 4.0  & 9.494 $\pm$ 0.003 & 25.4 $\pm$ 3.3 \\
4.0 - 6.0  & 9.466 $\pm$ 0.008 & 21.8 $\pm$ 3.4 \\
6.0 - 10.0 & 9.515 $\pm$ 0.031 & 12.8 $\pm$ 2.7 \\
\multicolumn{3}{c}{0-12\%} \\ 
2.5 - 3.0   & 23.531 $\pm$ 0.001 & 5.5 $\pm$ 1.5 \\
3.0 - 4.0   & 23.469 $\pm$ 0.002 & 5.4 $\pm$ 1.6 \\
4.0 - 6.0   & 23.395 $\pm$ 0.005 & 4.9 $\pm$ 1.5 \\
6.0 - 10.0  & 23.365 $\pm$ 0.021 & 3.0 $\pm$ 1.1 \\
\hline
\end{tabular}

\caption{The $v_2$ and the normalization values used for the
background subtraction in Fig.~\ref{fig:dphi_cent}.
}\label{tab:fig1}
\end{minipage}\hfill%
\begin{minipage}{\columnwidth}
\begin{tabular}{ccc}
\pTtrig\ & $B_{|\Delta\eta|<2.0}$ & $\langle v^{assoc}_2\rangle\langle v^{trig}_2\rangle $
	  \\ 
( GeV/$c$ ) &  & ( $10^{-3}$ ) \\
\hline
\multicolumn{3}{c}{0.5 \pTab 1.0 \GeVc} \\ 
2.5 - 3.0  &  59.833 $\pm$ 0.002 & 2.9 $\pm$ 0.8 \\
3.0 - 4.0  &  59.638 $\pm$ 0.003 & 2.8 $\pm$ 0.8 \\
4.0 - 6.0  &  59.366 $\pm$ 0.009 & 2.6 $\pm$ 0.8 \\
6.0 - 10.0 & 59.235 $\pm$ 0.034 & 1.6 $\pm$ 0.6 \\
\multicolumn{3}{c}{1.0 \pTab 1.5 \GeVc} \\ 
2.5 - 3.0  & 17.286 $\pm$ 0.001 & 5.0 $\pm$ 0.1 \\
3.0 - 4.0  & 17.238 $\pm$ 0.002 & 4.9 $\pm$ 0.1 \\
4.0 - 6.0  & 17.182 $\pm$ 0.005 & 4.4 $\pm$ 0.1 \\ 
6.0 - 10.0 & 17.154 $\pm$ 0.018 & 2.7 $\pm$ 0.1 \\
\multicolumn{3}{c}{1.5 \pTab 2.5 \GeVc} \\ 
2.5 - 3.0  & 6.245 $\pm$ 0.001 & 6.8 $\pm$ 1.9 \\
3.0 - 4.0  & 6.230 $\pm$ 0.001 & 6.6 $\pm$ 2.0 \\ 
4.0 - 6.0  &  6.213 $\pm$ 0.003 &  6.0 $\pm$ 1.9 \\ 
6.0 - 10.0 & 6.211 $\pm$ 0.011 & 3.7 $\pm$ 1.4 \\ 
\multicolumn{3}{c}{2.5 \pTab 4.0 \GeVc} \\ 
2.5 - 3.0  & 0.5210 $\pm$ 0.0002 & 8.1 $\pm$ 2.4 \\
3.0 - 4.0  & 0.5212 $\pm$ 0.0003 & 8.0 $\pm$ 2.5 \\
4.0 - 6.0  &  0.5207 $\pm$ 0.0008 & 7.2 $\pm$ 2.4 \\
6.0 - 10.0 & 0.5199 $\pm$ 0.0030 & 4.5 $\pm$ 1.7 \\
\hline
\end{tabular}
\caption{
  The $v_2$ values and the normalization used for the
  background subtraction in Fig.~\ref{fig:dphi_pttrig}. }\label{tab:fig2}
\end{minipage}
\end{table*}

The background normalization ($B$) and the elliptic
flow $\langle v_2^{assoc} \rangle \langle v_2^{trig} \rangle$ that were
used to subtract the background in Figs \ref{fig:dphi_cent} and
\ref{fig:dphi_pttrig} are reported in Tables \ref{tab:fig1} and \ref{tab:fig2}.

\end{document}